# Numerical Integration of stochastic differential equations

Riccardo MANNELLA
*Dipartimento di Fisica and*
*Istituto Nazionale Fisica della Materia, UdR Pisa*
*Piazza Torricelli 2, 56100 Pisa, Italy*
*email: mannella@difi.unipi.it*

The aim of this lecture is to study numerical algorithms for the integration of stochastic differential equations. I will derive an algorithm which exactly integrates the SDE, using a generalization of a Taylor series in the presence of stochastic forces. Given the complexity, we will find that this algorithm, although "exact", it is not particularly fast or, in general, the most convenient in a digital simulation, although it is a useful benchmark to test other algorithms, and I will try to improve it. I will discuss the features of different algorithms, both in terms of accuracy in a deterministic sense, and also in statistical terms, i.e. how well the algorithm is able to reproduce, for instance, the correct equilibrium distribution. I will then briefly introduce algorithms to integrate stochastic differential equations which are driven by correlated noise. The lecture will close with the discussion of a few algorithms for the special case of a two dimensional system in a potential, subject to damping and noise. The interest for this particular case is due to the importance and the interest in algorithms which can integrate, for instance, particles in the liquid state.

## 1  Basic algorithm and relation to Fokker-Planck equations

### 1.1  Introduction

A very simple and straightforward algorithm to numerically integrate one dimensional differential equations driven by a single stochastic force (external to the system) was introduced by Rao in [1]. Note that some of the technical terms used in this introduction will become clear in the rest of the lecture. The algorithm is really a Taylor expansion, or, more precisely, a one step collocation scheme. It is possible to use other integration schemes (for a review of some widely used algorithms see Mannella in [2]; more recent papers are references [3–11]). To integrate stochastic differential equations (SDE), we can basically have integration schemes based on predictor correctors [12–19] and schemes based on Runge-Kutta [20–25,4,5]. Interesting material can also be found in the papers by Rümelin [26] and Riggs Jr. [27,28] where, beside Runge-Kutta based approaches, other integration methods are presented and some some attention is devoted to the accuracy of the different schemes. Finally, other particular schemes, not immediately connected to the two classes just introduced are the ones proposed in [10,11]. Note that I will only deal with temporal noise: for SDE driven by temporal and spatial noise, the interested reader could refer to, for instance, [29] and references therein.

In the literature it is possible to find algorithms which share the same basic idea with the one I will discuss here: see for instance [30–36]. Yet a different approach is the one followed by Fox [37], where the deterministic force appearing in the sde is integrated via a Runge-Kutta whereas the stochastic force is integrated using a Taylor expansion [a]. The algorithms proposed in references [4,5] extend the Fox approach to an implicit Runge-Kutta treatment of also the stochastic force.

That are some reasons which make one step collocation schemes slightly preferable to predictor correctors and Runge-Kutta approaches. Predictor correctors of high orders, which by definition imply continuous and differentiable functions, seem to be of difficult or risky application to SDE which, by definition, have continuous but non-differentiable solutions. In practice, the algorithm one derives yields a trajectory which is indeed non-differentiable, but the derivation of the schemes seem to be rather empirical. We should add here that at low order, when it is possible to make a comparison, typical predictor correctors coincide with schemes based on Taylor expansions. As far as Runge-Kutta are concerned, we have similar problems to the ones mentioned for predictor correctors, with the added complication that schemes are known in the literature which in principle should coincide but in practice differ, the differences being due to different ways of expanding the equilibrium distribution of the stochastic force. There is however a more fundamental reason to prefer one-step collocation schemes. In this approach, for relatively small orders of the schemes, it is possible to simply evaluate the different stochastic integrals which appear in the SDE, with the bonus that the algorithm can be straightforwardly generalized to cover cases for which the stochastic force is non-white.

I will always use the Stratonovich calculus [40] to integrate the stochastic integrals which will be needed at the various stages. After the derivation of the algorithm, a section will be devoted to explain the details of using a given calculus and the relation between a stochastic differential equation and a Fokker-Planck equation.

### 1.2  Derivation of the algorithm

The most general SDE in which the stochastic force is linear can be written in the form

$$\dot{x}_i(t) = f_i(x_i(t), t) + g_i(x_i(t), t)\xi(t). \tag{1}$$

---

[a] According to private communications from R.F. Fox, the Runge-Kutta is normally a fourth order RK [38, 39].

I will always assume that $\xi(t)$ is a Gaussian variable. To describe it, then, I will need only the first two moments. For simplicity, at least in this section, these moments will always be given by (the symbol $\langle \ldots \rangle$ will denote averages over the noise realizations)

$$\langle \xi(t) \rangle = 0 \tag{2a}$$

$$\langle \xi(t)\xi(t') \rangle = \delta(t - t'), \tag{2b}$$

where $\delta(t)$ is the usual Dirac delta. Note that Eq. 2a also defines the spectral distribution of the stochastic force: the spectral distribution of the stochastic force is given by the Fourier trasform of the two-time correlation function. For this case, the two-time correlation function is a Dirac delta, hence the spectral distribution will be trivially flat (the so called white noise). The derivation of the basic algorithm is much more simple in this case. For the moment, I will assume that $f_i$ and $g_i$ are autonomous, i.e. they are not explicit functions of time.

Expand via the Taylor formula the functions in Eq. 1. It is possible to write (I will assume a summation over repeated indices)

$$f_i(x_j(t)) = f_i(x_j(0)) + \partial_j f_i(x_k(0))(x_j(t) - x_j(0)) + \ldots \tag{3a}$$

$$g_i(x_j(t)) = g_i(x_j(0)) + \partial_j g_i(x_k(0))(x_j(t) - x_j(0)) + \ldots, \tag{3b}$$

plus derivatives of higher orders. I can also rewrite the above equation as

$$f_i^t = f_i^0 + f_{i,j}^0 \delta x_j(t) + \frac{1}{2} f_{i,jk}^0 \delta x_j(t) \delta x_k(t) + \ldots \tag{4a}$$

$$\begin{aligned} g_i^t = {} & g_i^0 + g_{i,j}^0 \delta x_j(t) + \frac{1}{2} g_{i,jk}^0 \delta x_j(t) \delta x_k(t) \\ & + \frac{1}{3!} g_{i,jkl}^0 \delta x_j(t) \delta x_k(t) \delta x_l(t) + \ldots, \end{aligned} \tag{4b}$$

where, for instance, $g_{i,jk}^s \equiv \frac{\partial^2}{\partial x_j \partial x_k} g_i(x_l(t = s))$ and so on: the symbol $\delta x_i(t)$ means $x_i(t) - x_i(0)$.

Eq. 1 can be integrated formally writing

$$x_i(t) - x_i(0) = \int_0^t \dot{x}_i(s) ds = \int_0^t f_i(x_i(s)) ds + \int_0^t g_i(x_i(s))\xi(s) ds. \tag{5}$$

The integration is formal because the rhs of Eq. 5 still depends explicitly on $x_i(t)$.

The idea is now to integrate Eq. 5 for times between zero and $h$, where $h$ is a (small) integration time step, then to substitute the Taylor expansions of $f_i^t$ and $g_i^t$ in the r.h.s. of equation 5. Treating $h$ as an expansion parameter, it will be possibile to derive higher orders approximations for $x_i(h) - x_i(0)$, approximations which will be substituted back in the Taylor expansions of $f_i^t$ and $g_i^t$, generating terms of higher orders for $x_i(h) - x_i(0)$ and so on. A symbol widely used in the following will be $\delta x_i^\alpha(h)$ indicating the contribution to $x_i(h) - x_i(0)$ coming from the perturbation term of order $h^\alpha$.

It is now possible to write

$$\delta x_i(t) = \int_0^h (f_i^0 + f_{i,j}^0 \delta x_j(s) + \frac{1}{2} f_{i,jk}^0 \delta x_j(s) \delta x_k(s) + \ldots) ds$$
$$+ \int_0^h \xi(s)(g_i^0 + g_{i,j}^0 \delta x_j(s) + \frac{1}{2} g_{i,jk}^0 \delta x_j(s) \delta x_k(s)$$
$$+ \frac{1}{3!} g_{i,jkl}^0 \delta x_j(s) \delta x_k(s) \delta x_l(s) + \ldots) ds. \tag{6}$$

At the lowest possible order (i.e. $h^{1/2}$), obtained keeping only the term $g_i^0$ in the second integral appearing in equation 6, one has straightforwardly

$$\delta x_i^{1/2}(h) = g_i^0 Z_1(h) + o(h^1) \tag{7}$$

$$Z_1(h) \equiv \int_0^h \xi(s) ds = \sqrt{h} Y_1 \tag{8}$$

where $Y_1$ is a stochastic Gaussian variable with average zero and standard deviation one; Eq. 8 follows considering that $Z_1(h) \equiv \int \xi(t) ds$ is a linear combination of gaussian random variables, hence a gaussian variable. The moments of $Z_1(h)$ follow, having used Eq. 2a to compute the averages. It is very simple to show that $x_i^{1/2}(h)$ in Eq. 7 is of order $h^{1/2}$: it follows from the definition of $Z_1(h)$.

Now, let me substitute equation 7 into equation 6, keeping only the lowest order terms (i.e. $O(h^1)$):

$$\delta x_i^1(h) = \int_0^h f_i^0 ds + \int_0^h \xi(s) g_{i,k}^0 \delta x_k^{1/2}(s) ds = f_i^0 h + g_{i,k}^0 g_k^0 \int_0^h Z_1(s) \xi(s) ds$$
$$= f_i^0 h + \frac{1}{2} g_{i,k}^0 g_k^0 [Z_1(h)]^2 + o(h^{3/2}), \tag{9}$$

where the equality

$$\int_0^h Z_1(s) \xi(s) ds = \frac{1}{2} Z_1(h)^2$$

follows from the definition of $Z_1(h)$ (Eq. 8) and from the Stratonovich calculus for the stochastic integrals (see below).

The details of how the calculations follows can be found in Mannella [2], including a derivation of the various stochastic integrals which are needed, i.e.

$$Z_2(h) \equiv \int_0^h Z_1(s)ds = h^{\frac{3}{2}} \left\{ \frac{Y_1}{2} + \frac{Y_2}{2\sqrt{3}} \right\} \tag{10}$$

and

$$Z_3(h) \equiv \int_0^h [Z_1(s)]^2 ds \approx \frac{h^2}{3} \left\{ Y_1^2 + Y_3 + \frac{1}{2} \right\}, \tag{11}$$

where $Y_2$ and $Y_3$ are two uncorrelated gaussian deviates with average zero and standard deviation one.

The Taylor expansion around $t = 0$ reads (see Mannella in [2] for the intermediate algebra)

$$\begin{aligned}
x_i(h) = {} & x_i(0) + g_i^0 Z_1(h) + f_i^0 h + \frac{1}{2} g_{i,k}^0 g_k [Z_1(h)]^2 + Z_2(h) \left\{ f_{i,k}^0 g_k^0 - g_{i,k}^0 f_k^0 \right\} \\
& + h Z_1(h) g_{i,k}^0 f_k^0 + \frac{1}{3!} [Z_1(h)]^3 \left\{ g_{i,jk}^0 g_j^0 g_k^0 + g_{i,k}^0 g_{k,j}^0 g_j^0 \right\} \\
& + \frac{1}{2} f_{i,jk} g_j^0 g_k^0 Z_3(h) + g_{i,k}^0 \left\{ f_{k,j}^0 g_j^0 - g_{k,j}^0 f_j^0 \right\} \left\{ Z_1(h) Z_2(h) - Z_3(h) \right\} \\
& + \frac{1}{2} f_{i,k}^0 \left\{ f_k^0 h^2 + g_{k,j}^0 g_j^0 Z_3(h) \right\} + \frac{1}{4!} g_{i,k}^0 \left\{ g_{k,n}^0 g_{n,j}^0 g_j^0 + g_{k,jn}^0 g_j^0 g_n^0 \right\} [Z_1(h)]^4 \\
& + g_{i,k}^0 g_{k,j}^0 f_j^0 \left\{ \frac{h}{2} [Z_1(h)]^2 - \frac{1}{2} Z_3(h) \right\} + \frac{1}{4} g_{i,jk}^0 g_j^0 f_k^0 \left\{ h [Z_1(h)]^2 - Z_3(h) \right\} \\
& + \frac{1}{16} g_{i,jk}^0 g_{k,n}^0 g_n^0 g_j^0 [Z_1(h)]^3 + \frac{1}{4!} g_{i,jkn}^0 g_j^0 g_k^0 g_n^0 [Z_1(h)]^4 + o(h^{5/2}) \tag{12}
\end{aligned}$$

which is the final expression. It will be will shown further down in some selected examples that the algorithm of Eq. 12 is good enough for a general purpose integration, and that higher orders in $h$ are not really needed. On the other hand, to compute terms at (even) next perturbation order would be a very difficult task, due to the appearance of non-Gaussian stochastic terms (for instance, like $Z_3$ above) for which the appropriate statistics can only be approximated.

It is very simple to generalize the basic algorithm to take into account non-autonomous functions. It is enough to add to Eqs. 4 the terms due to the partial derivatives with respect to time of $f_i$ and of $g_i$, i.e. the terms $\dot{f}_i^0 t$, $\dot{g}_i^0 t$ and $\dot{g}_{i,j}^0 (x_j(t) - x_j(0)) t$ (higher order terms will not contribute at the perturbation order considered in Eq. 12). In practice

$$\delta x_i^{3/2}(h)' = \dot{g}_i^0 (hZ_1(h) - Z_2(h)) \tag{13a}$$

$$\delta x_i^2(h)' = \frac{h^2}{2} \dot{f}_i^0 + \frac{1}{2} g_k^0 \dot{g}_{i,k}^0 \left\{ h[Z_1(h)]^2 - Z_3(h) \right\} \tag{13b}$$

$$+ \frac{1}{2} g_{i,k}^0 \dot{g}_k^0 \left\{ h[Z_1(h)]^2 + Z_3(h) - 2Z_1(h)Z_2(h) \right\} \tag{13c}$$

are all the necessary corrections to Eq. 12 of order $O(h^2)$.

Note that the algorithm was derived under the assumption of only one stochastic forcing: the case of a multidimensional stochastic forcing is much more complex and we refer the reader to Mannella in [2] for further details.

### 1.3 Fokker-Planck equations, the Îto Stratonovich controversy and numerical simulations

As strange as it may sound, the Îto Stratonovich controversy is anything but a controversy [41]. For some time, within the stochastic physics community, there has been a debate on which should be the correct prescription to obtain the Fokker-Planck operator corresponding to a given Langevin equation. Let me clarify the problem with an example.

Suppose I have the Langevin equation

$$\dot{x} = f(x) + g(x)\xi(t) \tag{14}$$

where $\xi(t)$ is as usual a Gaussian variable with moments

$$\langle \xi(t) \rangle = 0 \tag{15}$$

$$\langle \xi(t)\xi(s) \rangle = 2D\delta(t - s) \tag{16}$$

it is in general possible to write an infinite number of different Fokker-Planck equations (the differential operator driving the probability distribution of the variable $x$). The most common prescriptions found in the literature are that due to Îto [42], and that due to Stratonovich [40]. In particular, if $P(x,t)$ is the probability distribution of $x$ as function of $t$ under the flux given by Eq. 14, we obtain (I (S) will refer to Îto (Stratonovich))

$$\frac{\partial}{\partial t} P(x,t) = \frac{\partial}{\partial x} \left\{ -f(x) + \frac{1}{2} \frac{\partial}{\partial x} [g(x)]^2 \right\} P(x,t) \quad \text{(I)} \tag{17}$$

$$\frac{\partial}{\partial t} P(x,t) = \frac{\partial}{\partial x} \left\{ -f(x) + \frac{1}{2} g(x) g'(x) + \frac{1}{2} \frac{\partial}{\partial x} [g(x)]^2 \right\} P(x,t) \quad \text{(S)}. \tag{18}$$

How can it be possible that given the same differential equation (Eq. 14) we should have two (better, infinite, as we will see) different differential operators for the evolution of the probability distribution, if (note the structure of Eq. 14) $g(x)$ depends explicitly on $x$? The solution is that Eq. 14 does not make sense *per se*. It is a fact that the stochastic process $\xi(t)$ in Eq. 14 is such that the quantity $\dot{x}$ is "kicked" by the noise, with the result that its integral ($x(t)$) is not continuous: in practice, it is impossible to evaluate the term $g(x)\xi(t)$ because it is unknown for which $x$ $g(x)$ should be computed.

The problem has to do with the intrinsic limits of the Riemann-Stieljes integral, in the presence of stochastic forces [43, 44]. Let me suppose that I have two functions $f(t)$ and $g(t)$, defined in the interval $a \leq t \leq b$. For each partition $P : a \leq t_0 \leq t_1 \leq \ldots \leq t_n = b$ I can build

$$S_p = \sum_{i=1}^{n} f(\xi_i) \left[g(t_i) - g(t_{i-1})\right], \tag{19}$$

where $t_{i-1} \leq \xi_i \leq t_i$. Define $|P| = \max_{1 \leq i \leq n}(t_i - t_{i-1})$. If it exists, and it is finite, the limit

$$\lim_{|P| \to 0} S_p = S, \tag{20}$$

I will term $S$ as the *Stieltjes integral* of $f(t)$ with respect to $g(t)$. I will symbolically write it as

$$S = \int_a^b f(t) \, dg(t). \tag{21}$$

Now, if I have [44] that $g(t)$ is the difference between two functions which are finite, monotonic and non-decreasing, and also that $f(t)$ is continuous, then it follows that the integral $S$ exists. Also, if $\int_a^b f(t) \, dg(t)$ exists, then the integral $\int_a^b g(t) \, df(t)$ exists too, and in particular I have

$$\int_a^b f(t) \, dg(t) = f(b)g(b) - f(a)g(a) - \int_a^b g(t) \, df(t). \tag{22}$$

Finally, if $g(t)$ is differentiable and $g'(t)$ and $f(t)$ are integrable, I have also that

$$\int_a^b f(t) \, dg(t) = \int_a^b f(t) \, g'(t) \, dt. \tag{23}$$

Let me now introduce a new quantity, $W_t$, known as *Wiener process*. $W_t$ is defined (see Eq. 14) as

$$W_t \equiv \int_0^t \xi(s) \, ds, \tag{24}$$

or, symbolically but formally less accurately,

$$dW_t \equiv \xi(t)\, dt. \tag{25}$$

Eq. 23 implies that if I had to evaluate the integral $\int_0^t g(s)\,\xi(s)\,ds$, I could instead evaluate the integral

$$\int_0^t g(s)\,\xi(s)\,ds \equiv \int_0^t g(s)\,dW_s = g(t)W_t - \int_0^t W_s\, g'(s)\,ds \tag{26}$$

which obviously is defined *only if $g(s)$ does not depend on $W_t$*, otherwise the function $g'(t)$ is not defined.

The solution of this problem is to extend the original definition of Riemann-Stieltjes integral to include the case of stochastic functions. Let me first introduce the characteristic function of the interval $[a,b]$, which I will call $\chi_{[a,b]}(t)$, defined as

$$\chi_{[a,b]}(t) = \begin{cases} 1 & \text{if } a \le t \le b \\ 0 & \text{otherwise} \end{cases} \tag{27}$$

Let $f(t)$ be a *step function* in the interval $[a,b]$, i.e., if I have the partition $a \le t_0 < t_1 < \ldots < t_n = b$,

$$f(t) = \sum_{i=0}^{n-1} f(t_i)\chi_{[t_i,t_{i+1}]}(t). \tag{28}$$

I will define (Îto)

$$\text{(I)} \quad \int_a^b f(t)\,dW_t \equiv \sum_{i=0}^{n-1} f(t_i)\left[W_{t_{i+1}} - W_{t_i}\right]. \tag{29}$$

Eq. 29 is the definition of integration according to the Îto prescription. We note that the function $f(t)$ could in principle be a stochastic function, given that in the Îto prescription only the values it assumes at the times $t_i$ matter. Physically, we could interpret Eq. 29 saying that to evaluate the stochastic integral I must take the value the function $f(t)$ assumes *immediately before* the stochastic term is "applied", to overcome the problems with the discontinuity of the variable $t$. From Eq. 29 I can obtain the following identity [43, 44] (integration by parts)

$$\text{(I)} \quad \int_a^b W_s\, dW_s = \frac{1}{2}\left(W_b^2 - W_a^2\right) - \frac{1}{2}\sum_{n=0}^{\infty}(dW_n)^2 = \frac{1}{2}\left(W_b^2 - W_a^2\right) - \frac{1}{2}(b-a). \tag{30}$$

It should be clear that the definition introduced in Eq. 29 is somehow arbitrary. Strictly from a mathematical point of view, the Îto integral has interesting properties (see [43]) which makes it formally most attractive. From a physical point of view, on the other hand it is clear that Eq. 29 is not symmetric with respect to the variable $t$ (somehow it "points towards the future"). Also, Eq. 30 show that the usual rules for parts integration do not apply to the Îto definition. A different prescription, which is symmetric with respect to the variable $T$ and for which the usual part integration rules apply is the one introduced by Stratonovich [40]. Using the same hypothesis used for Eq. 29 we can define

$$(S) \quad \int_a^b f(t)\, dW_t \equiv \sum_{i=0}^{n-1} \frac{f(t_i) + f(t_{i+1})}{2} \left[W_{t_{i+1}} - W_{t_i}\right]. \quad (31)$$

Clearly, Eq. 31 is symmetric with respect to the variable $t$; also, we have that

$$(S) \quad \int_a^b W_s\, dW_s = \frac{1}{2}\left(W_b^2 - W_a^2\right). \quad (32)$$

In general, it is even possible to think of some "mixed" prescription, which could be defined as

$$(M) \quad \int_a^b f(t)\, dW_t \equiv \sum_{i=0}^{n-1} \left[(1 - \frac{M}{2})f(t_i) + \frac{M}{2}f(t_{i+1})\right] \left[W_{t_{i+1}} - W_{t_i}\right], \quad (33)$$

for which we have

$$(M) \quad \int_a^b W_s\, dW_s = \frac{1}{2}\left(W_b^2 - W_a^2\right) - \left(\frac{1}{2} - M\right)(b-a), \quad (34)$$

which yields the Îto (Stratonovich) prescription taking $M = 0$ $\left(\frac{1}{2}\right)$.

Typically a numerical algorithm (and in particular the one step collocation derived here, as mentioned before) uses heavily a standard rule of integration by parts for the evaluation of the stochastic integrals at different orders. This implies that the algorithm integrates according to the Stratonovich prescription. If it were necessary to integrate according to the $M$ prescription (remembering that for Îto we have $M = 0$), it would be enough to replace $f_i$ in Eq. 1 with $f_i + \left(M - \frac{1}{2}\right)\left(\frac{\partial}{\partial x_j}g_i\right)g_j$, assuming a summation of repeated indices. Let me stress that a numerical simulation *cannot*, under any circumstance, resolve the Îto Stratonovich controversy, because the actual form given to the integrating algorithm must be decided beforehand: and if a prescription $M$ is picked to write the algorithm, obviously the integration scheme will integrate according to $M$.

It is possible to say that the Stratonovich prescription corresponds to considering a stochastic process with a finite correlation time, where the limit of zero correlation time is taken once all the different integrals have been computed. This poses the following problem: in practice, what is the "correct" prescription to integrate Eq. 14? A possible answer is the following: if the "internal" fluctuations of the variable $x$ are faster than the fluctuations of the variable $\xi(t)$, then it is reasonable that the correct prescription will be more Îto like. In the opposite limit, the Stratonovich prescription will be the norm. This has been shown to be the correct picture in some analogue simulations [45], where, electronically simulating the Langevin equation

$$\dot{x} = f(x) + g(x)\xi(s) + \eta(t) \qquad (35)$$

(where $\xi(t)$ and $\eta(t)$ are two uncorrelated stochastic processes) it is possible to show that the equilibrium distribution of the variable $x$ will move from the one characteristic of Îto prescription to the one typical of the Stratonovich prescription simply changing the relative (short) correlation times of the two stochastic processes.

## 1.4  Improving the basic algorithm

In principle it is possible to improve the basic algorithm combining it with a suitable number of predictor-correctors [38]. The first attempt in this direction, following the approach we have shown for the derivation of the stochastic variables in the basic algorithm, it is probably the one by Blum [46]. He introduced the so called Heun scheme, which has been subsequently used also by Rümelin [26]. In practice the method consists of an Adams Bashforth (AB) predictor of order zero corrected via an Adams Moulton (AM) corrector of order one, where the basic algorithm at first order is used to evaluate the forces in both stages (see Eqs. 7 and 9). We would like to point out that the introduction of an AM corrector, which weights equally the stochastic forcing at the beginning and at the end of the integration time step, will automatically implement the Stratonovich calculus for the SDE (see previous chapter for more details). It would not be necessary, then, to introduce the term $\frac{1}{2}g^0_{i,j}g^0_j[Z_1(h)]^2$ (see Eq. 9), which in the basic algorithm made sure that the calculus employed was really the Stratonovich one.

Let me define AB($n$) (AM($n$)) as an Adams Bashforth predictor (Adams Moulton corrector) of order $n$. More in details, having to integrate the SDE

$$\dot{x} = f(x) + g(x)\xi \qquad (36)$$

$$\langle \xi(t) \rangle = 0 \tag{37}$$

$$\langle \xi(t)\xi(s) \rangle = \delta(t-s), \tag{38}$$

with $\xi$ Gaussian, according to the Heun algorithm at each integration step one will have to compute

1. $\tilde{x}(h) = x(0) + f(x(0)) + g(x(0))Z_1(h)$

2. $x(h) = x(0) + \frac{1}{2}\left[f(x(0)) + f(\tilde{x}(h))\right] + \frac{1}{2}\left[g(x(0)) + g(\tilde{x}(h))\right]Z_1(h).$ (39)

and go through the loop again for the following step of integration.

Independently, Mannella and Palleschi [47] used an algorithm which consists of an AB predictor of zero-th order, corrected via an AM corrector still of zero-th order, but using, as the building block for the forces, the basic algorithm at order $h^{3/2}$. In this case, however, it is important to note that the corrector requires the evaluation of the stochastic integral corresponding to $Z_2(h)$ around $h$. The corresponding quantity is evaluated in [48]. However, this algorithm is here quoted only for the sake of completeness, because from the point of view of reconstructing the correct equilibrium distribution it does not do better than simpler and faster algorithms.

## 2  Behavior of the different integration schemes

As mentioned in the previous pages, when considering the integration of SDE, one has two different aspects: the accuracy of the integration of the "deterministic" part of the SDE and how well represented are the statistical quantities associated with the given SDE.

In the following I will use two tools to study the different schemes. I will use, for the deterministic drift, the standard tool normally employed to work out the coefficients in a predictor corrector scheme (it will be clear in the following how the technique works). For the statistical properties, I will work out the actual equilibrium distribution obtained in the numerical integration: and from a comparison with the exact quantities we will easily judge pros and cons of the various schemes. This is achieved deriving the Fokker-Planck equation obtained in the numerical scheme. Suppose that the discrete integration scheme reads

$$x_i(t+h) = x_i(t) + F_i(x,t) \tag{40}$$

(see the structure of Eq. 12), then the following partial differential equation follows ($P$ is the probability distribution)

$$P(t+h) - P(t) = \sum_{n=1}^{\infty} \sum_{x_1\ldots x_n} \frac{\partial}{\partial x_1} \cdots \frac{\partial}{\partial x_n} K_{1\ldots n} P(t) \tag{41}$$

where
$$K_{1\ldots n} \equiv (-)^n \frac{1}{n!} \langle F_1 \ldots F_n \rangle_{noise}. \tag{42}$$

This equation is obtained [49] from the evolution equation

$$P(x_i, t+h) = \langle \int dy_1 \ldots dy_n P(y_i, t) \Pi_{y_i} \delta(x_i(t+h) - y_i(t) - F_i(y, t)) \rangle_{noise} \tag{43}$$

by Taylor expansion and average over the noise realization.

The r.h.s. in Eq. 41 will turn out to be always of the form

$$\text{r.h.s.} = \sum_{n=1}^{\infty} h^n s_n, \tag{44}$$

which means that we can divide both members of Eq. 41 by $h$. In the limit $h \to 0$ the l.h.s. of Eq. 41 will give the derivative of $P(t)$ with respect to time. However, depending on how the various powers of $h$ will vanish in the r.h.s. of Eq. 41 , we will infer which integration scheme is preferable at finite integration time steps.

I will look at the following cases:

**Heun scheme:** The Heun scheme was described in Eq. 39

**Simple Euler:** The simple Euler corresponds to the algorithm given by Eqs. 7 and 9.

**Exact propagator:** I will define as the exact propagator algorithm an algorithm where I integrate exactly (i.e., with a very high order integration time step) the *deterministic* part of the equations of motion, and then the contribution given by Eq. 7 is added to the deterministic flow.

**Algorithm $o(h^2)$:** It will be the integration scheme given by Eq. 12, dropping the $O(h^2)$ terms.

**$Z_3$-less:** It will be the algorithm given by Eq. 12, but without the terms containing $Z_3$.

**Full algorithm:** The algorithm given by Eq. 12. I will also refer to this case as to the algorithm $O(h^2)$.

We are now ready to compare the different algorithms.

## 2.1 Deterministic behavior of the different integration schemes

The evaluation of the error associated with a given integration scheme parallels the derivation of a predictor corrector scheme. To understand how the thing is done, I will derive the AM corrector to first order. Starting from the equation

$$\dot{x} = f(x), \tag{45}$$

the first order AM reads

$$x(t+h) = x(t) + h(a_1 f(x(t+h)) + a_0 f(x(t))). \tag{46}$$

We need to evaluate (to find the optimal values) for the parameters $a_0$ and $a_1$. The idea is to write

$$x(t+nh) = (t+nh)^m \tag{47}$$

where $m$ is by one unity larger than the unknowns one has to optimize: in this case, $m = 3$.

Inserting Eq. 47 in Eq. 46, noting that $f(x(t+nh)) = \dot{x}(t+nh)$, I have

$$(t+nh)^3 = t^3 + h(3a_1(t+h)^2 + 3a_0 t^2) \tag{48}$$

from which it follows

$$t^3 + 3t^2 h + 3th^2 + h^3 = t^3 + t^2 h(3a_1 + 3a_0) + 6th^2 a_1 + 3h^3 a_1. \tag{49}$$

Equating now the terms which have the same power of $t$, we immediately find the standard result, $a_1 = a_0 = 1/2$. However, it is evident that after all algebra, we are left not with zero, but with the quantity

$$h^3 = 3/2 h^3 \tag{50}$$

which implies that the intrinsic error associated with this method will be $O(h^3/2)$.

For a simple minded (Euler) integration one has

$$(t+h)^2 = t^2 + 2ht \tag{51}$$

which leads to an intrinsic error which is $O(h^2)$.

The algorithm of Eq. 12, if restricted to the deterministic part, implies that

$$x(t+h) = x(t) + hf(x(t)) + \frac{h^2}{2} f(x(t)) \frac{df(x(t))}{dx(t)}. \tag{52}$$

Noting that $\dot{x} = f(x(t))$ and that $\ddot{x} = \dot{x}f'(x(t)) = f(x(t))f'(x(t))$, it follows that in this case

$$x(t+h) = x(t) + h\dot{x}(t) + h^2\ddot{x}/2 \tag{53}$$

and, using $x(t) = t^3$, I have

$$(t+h)^3 = t^3 + 3ht^2 + 3h^2t \tag{54}$$

which leads to an intrinsic error which is $O(h^3)$. This error is of the same order of the AM corrector I examined before, although it is somehow larger (note that in the AM corrector example we had a factor $1/2$ in front of $h^3$). We should expect, hence, that the AM corrector is slightly more accurate than the algorithm of Eq. 12 in the limit of very small noise intensities.

I will now verify these results in a simple stochastic system. The system considered in the numerical simulations is a model of dymer in the presence of thermal fluctuations [50–52]: it is the semiclassical approximation of a nonlinear Schrödinger equation which can be cast in the form of a classical Langevin equation. This dymer, incidentally, is basically equivalent to a spin in the presence of a magnetic field along the $z$ axis and of a fluctuating magnetic field on the $x$ axis. The SDE one writes is in the form [51]

$$\dot{p} = 2Vp \tag{55a}$$

$$\dot{q} = -2Vp - \chi rp - \chi rz \tag{55b}$$

$$\dot{r} = \chi qp + \chi qz \tag{55c}$$

$$\dot{z} = -\Gamma z + \xi(t) - 2Vq \tag{55d}$$

$$\langle \xi(t) \rangle = 0$$

$$\langle \xi(t)\xi(s) \rangle = 2D\delta(t-s),$$

where the different $p, q$ are $r$ are suitable linear combinations of the elements of the dymer density matrix, and the stochastic force $\xi(t)$, which is assumed to have a Gaussian statistics, represents the interaction between the dymer and a thermal bath with finite temperature. Although the last equation in 55 is reminiscent of a colored noise evolution, the presence of the reaction term $-2Vq$ in the equation for $\dot{z}$ implies that colored noise algorithms cannot be used. In practice it would be straightforward to modify these algorithms to cover this case, given the linearity of the equation for $\dot{z}$, but here I really want to learn something about the different "deterministic" schemes, and I will then simply use standard white noise algorithms.

Apart from the wide relevance of Eq. 55 in the physical sciences (the onset of possible localized states, for instance, could have consequences on the onset of Davydov solitons in molecular chains), their importance here is that they admit an exact integral of motion: in fact, despite the nonlinearity, by inspection it is possible to note that the quantity $p^2+q^2+r^2$ is an exact constant of motion (remembering that the equations are some kind of semiclassical approximation, it is possible to show that the constant equals one). For the numerical experiments presented further down (figure 1) the parameters chosen were $\chi = 1.25$, $\Gamma = 3$, $2V = 1$, starting from the initial condition $p = 1$, $q = r = 0$. This implies that for the subsequent evolution $p^2 + q^2 + r^2 = 1$. A test particle was then followed for times equal to 200, and the final value of the quantity $p^2 + q^2 + r^2 - 1$ was plotted as function of the integration time step, for different integration algorithms and for two different noise intensities $D$.

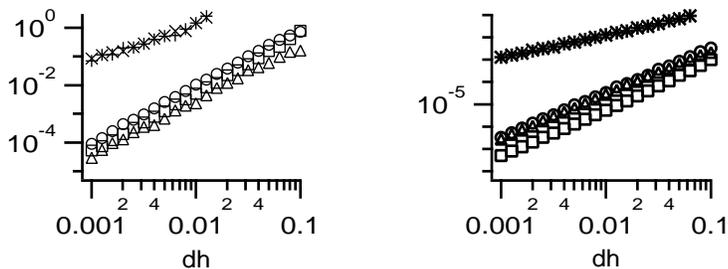

Figure 1: Algorithms for the dymer model, with $D = 1.0$ (left) and $D = 0.01$ (right). $\bigcirc$ exact propagator, $+$ Simple Euler, $\square$ Heun scheme, $\times$ Algorithm $o(h^2)$, $\triangle$ Algorithm $O(h^2)$.

For both intensities of the noise considered, the algorithms which are first order in the deterministic force are considerably less accurate in preserving the norm. Of the higher order algorithms, the Heun scheme does better for small noise intensities, whereas the full algorithm is better for large noise intensities. The algorithm which integrates exactly the deterministic force is better than the first order algorithms, but it is inferior to both the Heun scheme and the full algorithm.

### 2.2 Statistical behavior of the different integration schemes

The tool used to understand the statistical error associated with each integration scheme is Eq. 41, where I will insert the numerical scheme in the expression for $K_{1...n}$ and average over the noise variables.

I will also limit my investigation, without loss of generality, to one dimensional system driven by one additive noise. The general system I will look at

is in the form
$$\dot{x} = -V'(x) + f(t) \tag{56}$$
where $f(t)$ is a gaussian random process, with moments
$$\langle f(t) \rangle = 0 \quad \langle f(t)f(s) \rangle = 2D\delta(t-s). \tag{57}$$

Associated with the stochastic differential Eq. 56 we also have a Fokker-Planck equation, which reads
$$\frac{\partial P(x,t)}{\partial t} = -\frac{\partial}{\partial x}\left[V'(x) + D\frac{\partial}{\partial x}\right]P(x,t). \tag{58}$$

Note that this equation is in the form of a conservation equation
$$\frac{\partial P(x,t)}{\partial t} + \frac{\partial J(x,t)}{\partial x} = 0 \tag{59}$$

with the "probability current" $J(x,t)$ given by the equation
$$J(x,t) \equiv \left[V'(x) + D\frac{\partial}{\partial x}\right]P(x,t) \tag{60}$$

The equilibrium solution of Eq. 58 is readily obtained. First, given that we are talking of an equilibrium solution (I will also use mild hypotheses, like that the function $V(x)$ grows quickly enough for large $x$'s, so that a ground state for $P(x,t)$ is defined), I can say that $\frac{\partial P(x,t)}{\partial t}$ must vanish. This leaves me with the equation
$$-\frac{\partial}{\partial x}\left[V'(x) + D\frac{\partial}{\partial x}\right]P(x,t) = 0. \tag{61}$$

This equation implies that the divergence of the current must vanish: this implies that the current itself must be a constant. However, under the mild hypotheses mentioned above, the current must also vanish for $|x| \to \infty$, which implies that the current must vanish identically for any $x$ once the equilibrium solution has been reached.

This leads to the equation
$$\left[V'(x) + D\frac{\partial}{\partial x}\right]P(x,t) = 0 \tag{62}$$

which has the solution
$$P(x,t) = N\exp\{-V(x)/D\} \tag{63}$$

where $N$ is a normalization constant.

In general, integrating Eq. 56 with a discrete integration routine we will have an equilibrium solution which is not exactly given by Eq. 63, but rather by

$$P(x,t) = \tilde{N} \exp\left\{\left(-V(x) + \sum_{n=1} h^n S_n(x)\right)/D\right\}. \tag{64}$$

In the figures which follow I have simulated, using the different schemes, the stochastic dynamics of the systems

$$\dot{x} = -x - x^3 + f(t) \tag{65}$$

and

$$\dot{x} = x - x^3 + f(t), \tag{66}$$

and I will then compare the result of the simulations with the "wrong" equilibrium distribution which is the equilibrium distribution obtained using each integration scheme with a finite time step. In the simulations, $f(t)$ is a gaussian random variable as in Eq. 56. The comparison between theory and numerical simulations is shown in figure 2.

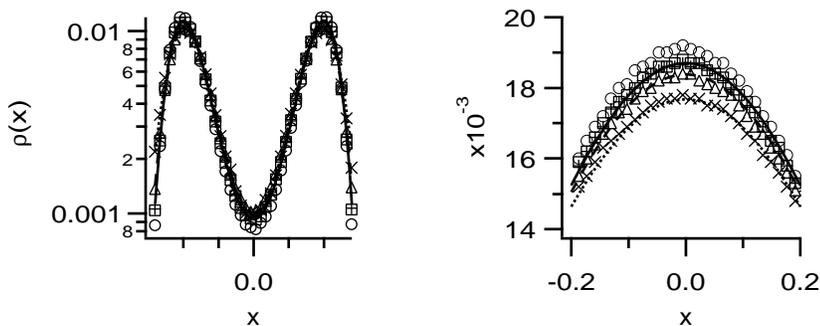

Figure 2: Equilibrium distributions obtained using different integration schemes, $D = 0.1, h = 0.1$. $\times$ and dotted line: exact propagator; $\triangle$ and dashed line, simple Euler; $+$, algorithm $O(h^2)$; $\square$, Heun algorithm; $\bigcirc$, $Z_3$-less; Solid line, exact equilibrium distribution. Left, Eq. 66; right, distribution near the maximum for Eq. 65.

2.3  *Simple Euler*

In this case, the integration scheme reads

$$x(t+h) = x(t) + hf(x(t)) + \sqrt{2Dh}\eta \tag{67}$$

and we need only to keep the lowest four terms in Eq. 42. Carrying out the necessary algebra, the equilibrium distribution reads

$$P(x,t) = \tilde{N} \exp\left\{\left[-V(x) + \frac{h}{2}\left(V'(x)^2/2 - DV''(x)\right)\right]/D\right\}. \qquad (68)$$

Note that the difference with respect to the true equilibrium distribution is order of $h$ in the exponent. It is clear from the figures that indeed the simulations carried out using a simple Euler scheme follow very closely the expected theoretical distribution.

2.4  Exact propagator

In this case, I need to first solve the equation

$$\dot{x} = f(x) \qquad (69)$$

over the time interval $[t, t+h]$, starting from $x = x(t)$; call $x(t+h)_{det}$ the solution at $t = t+h$. Then, the stochastic $x(t+h)$ is given by

$$x(t+h) = x(t+h)_{det} + \sqrt{2Dh}\eta. \qquad (70)$$

As above, it is possible to insert this prescription in the equation for the moments 42. After carrying out the necessary algebra, the equilibrium distribution reads

$$P(x,t) = \tilde{N} \exp\left\{\left[-V(x) + \frac{h}{2}\left(V'(x)^2 - DV''(x)\right)\right]/D\right\}. \qquad (71)$$

which again differs from the true equilibrium distribution by a term order of $h$ in the exponent. Incidentally, it is clear that at very small $D$ the equilibrium distribution obtained using an exact propagator is *worse* than the equilibrium distribution obtained using a simple Euler scheme, as it is clear both from the figure and by inspecting the corresponding equilibrium distribution.

This should already be a very important warning: when dealing with SDE, it is extremely delicate how higher orders are introduced; and if the orders in the approximation in the deterministic and in the stochastic part of the flow are not "balanced", when we increase the order of the algorithm we may actually make it worse!

## 2.5 Algorithm $o(h^2)$

In this case we find a result similar to the one found for the simple Euler case: the equilibrium distribution generated by this scheme is given by

$$P(x,t) = \tilde{N} \exp\left\{\left[-V(x) + \frac{h}{2}\left(V'(x)^2/2 - DV''(x)\right)\right]/D\right\}. \tag{72}$$

## 2.6 $Z_3$-less

In this case we find already an improvement with respect to the previous cases: the equilibrium distribution reads

$$P(x,t) = \tilde{N} \exp\left\{\left[-V(x) - \frac{h}{2}\left(DV''(x)\right)\right]/D\right\}. \tag{73}$$

In other words, the term proportional to $V'(x)^2$ disappears. This implies that, comparing the results for the Euler scheme, the exact propagator, the algorithm $o(h^2)$, and Eq. 12, the term $V'(x)^2$ is canceled both by the joint "action" of the term $Z_2$ and by the term proportional to $f_{i,k}f_k$ in Eq. 12. But it also means that dropping one of these two ingredients yield a much less accurate integration scheme. Although I will show below that both the Heun scheme and the full algorithm are actually better integration scheme, for very small $D$ the $Z_3$-less algorithm is a significant improvement with respect to the other algorithms seen so far.

## 2.7 Heun scheme

The algebra corresponding to this case is somehow more cumbersome, because now the correction to the true equilibrium distribution is order of $h^2$ in the exponent. Writing the equilibrium distribution in the form

$$P(x,t) = \tilde{N} \exp\left\{\left[-V(x) + h^2 F(x)\right]/D\right\}, \tag{74}$$

the function $F(x)$ satisfies the equation $(V \equiv V(x), F \equiv F(x))$

$$0 = 12D^2 F' + 2V'^5 - 4DV'^3 V'' + 33D^2 V' V''^2 + 21D^2 V'^2 V''' \\ 1 - 24D^3 V'' V''' - 2D^3 V' V^{(IV)} + 3D^4 V^{(V)}. \tag{75}$$

The importance of the Heun scheme, however, is that the correction is only order of $h^2$ in the exponent, which means that the equilibrium distribution it generates will in general be fairly close to the true one, as it is confirmed by the simulations.

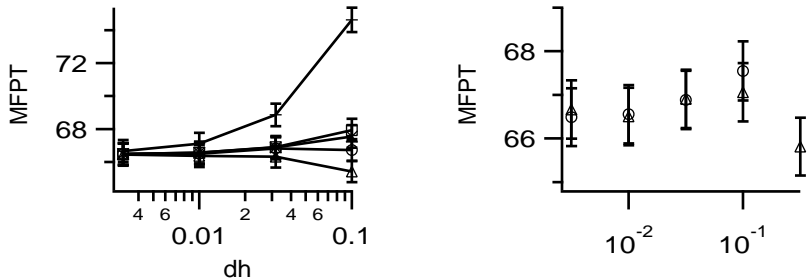

Figure 3: Left: MFPT with different algorithms, for $D = 0.1$, vs $dh$. $\bigcirc$, full algorithm; $\square$, exact propagator; $+$, algorithm $o(h^2)$; $\times$, Heun algorithm; $\triangle$, simple Euler. Right: Comparison between the MFPT obtained using the white noise Heun algorithm ($\triangle$), and the MFPT obtained using the coloured noise Heun algorithm, with $\tau = 10^{-4}$ ($\bigcirc$).

### 2.8 Full algorithm

As in the case of the Heun scheme, the algebra in this case is very cumbersome. The correction to the true equilibrium distribution is again order of $h^2$ in the exponent. Writing the equilibrium distribution in the form

$$P(x,t) = \tilde{N} \exp\left\{ \left[ -V(x) + h^2 F(x) \right] / D \right\}, \tag{76}$$

the function $F(x)$ satisfies the equation ($V \equiv V(x), F \equiv F(x)$)

$$\begin{aligned} 0 = {}& 36 D^2 F' + 6 V'^5 - 60 D V'^3 V'' + 96 D^2 V' V''^2 + 78 D^2 V'^2 V''' \\ & - 90 D^3 V'' V''' - 66 D^3 V' V^{(IV)} + 24 D^4 V^{(V)}. \end{aligned} \tag{77}$$

It is clear that the two best algorithms, in the sense that they generate an equilibrium distribution which is the closest to the "true" one, are the Heun scheme and the "full algorithm". It should be appreciated that even if the correction in the equilibrium distributions may look tiny (one could always think of making the integration time step smaller), the actual Fokker-Planck equation is different, and the difference in the equilibrium distribution is in an exponent. Hence, we may expect unpredictable results when things like the mean first passage time between two minima in a bistable potential are computed. What is worse, for fixed $h$ the correction depends on $D$, making it more difficult to compute, for instance, activation energies from the simulations.

To show that the calculation of the Mean First Passage Time (MFPT) is indeed strongly influenced by the algorithm chosen, I summarized the result of computing the mean first passage time between two minima in the potential $V(x) = -x^2/2 + x^4/4$, using different algorithms and integration time steps in figure 3. The noise intensity was always $D = 0.1$: note how the convergence to the "limiting" MFPT is much slower when I used algorithms which lead to an order $h$ correction in the equilibrium distribution (a similar conclusion could have been guessed from figure 3). I used 10000 averages for each point in figure, so that the statistical scattering is order of 1.0%. Overall, the full algorithm is the one which stays closer to the limiting MFPT as the integration time step is changed, with the Heun algorithm which is typically also very close. The expected theoretical value for the mean first passage time is 66.3.

I will finish this section with yet another example which should warn us about the care which must be exerted when higher order integration schemes are derived. One of the earliest, and most "transparent", derivation of integration schemes, based on Runge-Kutta, is the one in [53]. There a first algorithm is derived as an example (termed $2_O 2_S 1_G$ by the author, meaning a second order deterministic one, second order stochastic one and needing the generation of one gaussian deviate), and a second algorithm is then given in appendix ($3_O 3_S 2_G$, supposedly a better and more accurate algorithm), as a "production" algorithm.

Now, whereas the "example" algorithm of [53] is equivalent to the Heun scheme, and it shares all "good" properties of this algorithm, the $3_O 3_S 2_G$ scheme leads to the equilibrium distribution

$$P(x,t) = \tilde{N} \exp\left\{[-V(x) - h\left(0.145186 D V''(x)\right)]/D\right\}. \tag{78}$$

i.e. it generates an equilibrium distribution which differs by the correct one by an order $h$ in the exponent, and we should expect problems similar to the ones typically found in very low order algorithms. Also, it apparently does not do much better than the $Z_3$-less scheme, which requires one less calculation of the deterministic force for each integration step.

## 3 Higher order stochastic differential equations

### 3.1 One pole and two poles filters

An important role is played by noisy drivings which are not white, but with a spectral density which is Lorentzian. In particular, systems where the stochastic forcing is first passed through a one pole or a two poles filters. If $f(t)$ is the usual white noise gaussian process, and $y(t0$ is the noise driving the system of interest, $y(t)$ is described by the equations

$$\dot{y} = -\frac{1}{\tau} y + f(t) \tag{79}$$

$$\ddot{y} = -\gamma \dot{y} - \omega_0^2 y + f(t). \tag{80}$$

The linearity of these equations has been exploited in [47] and [54,55] to derive an algorithm which parallels the algorithm of Eq. 12, with suitable $Z_1$, and $Z_2$. Also in this case we can improve the basic algorithm using the Heun algorithm (Eq. 39). The important feature of these algorithms is that they are fully implicit for the integration of the noise: this implies that they will exactly reproduce the evolution of the filtered noise, irrespectively of the relative ratio between the integration time step and the intrinsic time scales of the filters. This is confirmed using the algorithm for a one pole filter to integrate a quasi white noise: in figure 3 I plotted the result, and we should note that for all $dh$ considered, $dh$ is larger than $\tau$.

*3.2  Some alternative algorithms*

For the specific case of exponentially correlated noise it is possible to find in the literature some alternative integration scheme. Here I will briefly review the one introduced in [7,9] and the one introduced by [10].

A very elegant and efficient scheme for exponentially correlated noise has been proposed in [7,9]. The idea is to start from

$$\dot{x} = f(x) + y \tag{81}$$

with $y$ gaussian and correlated as per

$$\langle y(t) y(s) \rangle = F(t-s).$$

Clearly, we have through a Fourier transform

$$\langle y(\omega) y(\omega') \rangle = G(\omega - \omega') = H(\omega) \delta(\omega - \omega').$$

Now, write (central limit theorem)

$$y(jh) = (1/\sqrt{h}) \sum_{i=0}^{N} a_i \cos(i \Delta \nu j h + \phi_i) \tag{82}$$

with $\phi_i$ uniformly distributed over $[0, 2\pi]$, $a_i^2 = H(i \Delta \nu) \Delta \nu$, and suitably chosen $N$ and $\Delta \nu$, such that the conditions

$$\Delta \nu \ll \nu_s$$

and

$$N \Delta \nu \gg \nu_s$$

are satisfied. It is then possible to integrate Eq. 81. The key point (see [7]) is that the sum appearing in Eq. 82 can be evaluated with a Fast Fourier Transform, which implies that to generate, say, $y(ih)$ for times from 0 up to $Nh$, only $N \log N$ operations are required. Although for one pole or two poles filters the direct simulation would be faster, for computationally expensive $H(\omega)$ this method is much preferable [8,7]. For vector and parallel machines, furthermore, the whole algorithm can be made to run very efficiently.

Another algorithm to integrate exponentially correlated noise is the one proposed in [10]. The idea is to replace the exponentially correlated noise by a superposition of independent random telegraph processes. Suppose we have $N$ random telegraph processes $\{s_i\}$, oscillating between the states $+1$ and $-1$, and such that

$$\langle s_i \rangle = 0 \qquad \langle s_i(t) s_j(0) \rangle = \delta_{ij} e^{-2a|t|}. \tag{83}$$

Introduce the variable $Y$, defined as

$$Y \equiv \frac{1}{2} \sum_{i=1}^{N} s_i, \tag{84}$$

which takes values between $-N/2$ to $N/2$ in steps of unity. If $M = Y + N/2$ is the number of $s_i = 1$ in the sum, then the variable $Y$ changes its state to $Y \to Y \pm 1$ with a probability $M/N$ for $Y \to Y-1$ and $1-M/N$ for $Y \to Y+1$.

It is possible to work out the statistical properties of the variable $Y$: one finds [10]

$$\langle Y \rangle = 0 \qquad \langle Y(t)Y(0) \rangle = (N/4) e^{-2a|t|}, \tag{85}$$

and, given that $Y$ is the sum of independent processes, the quantity $Y/N$ becomes a gaussian in the limit of large $N$. We have clearly a possible representation of an exponentially correlated noise. The apparent advantage with respect to other approaches is that, between different times at which the random telegraph signal switches, the noise is "constant", and one is left to integrate a *standard* differential equation. The method picks the time sequence $\{t_k\}$ at which the variable $Y$ switches using the prescription

$$t_{k+1} = t_k + \eta/(Na) \tag{86}$$

where $\eta$ is a random variable exponentially distributed with a cutoff equals to one: clearly, the larger $N$ (to achieve a better gaussianicity), the smaller the time steps generated.

Apart from the problems connected with the determination of the optimal $N$ (which is a difficult question, in practice, and can only perhaps be inferred in an indirect way, for instance checking for consistency of the results as $N$ is changed), it is not obvious at all that the method should yield faster algorithms or, more important, that it reproduces distributions which are close to the correct one. In fact, taking a concrete examples, I integrated with this algorithm the SDE corresponding to the usual bistable potential, for $\tau = 1.0$ and $D = 0.1$. I followed the stochastic trajectory for a time equals to $10^4$. The results are summarized in table 1, where I used $N = 21$ as suggested by the authors of [10]. At first sight, the present algorithms is better as I decrease the integration time step (although, as I have shown before, the Heun scheme works even for time steps as large as 0.1). The problem, however, is when I check the gaussianity of the noise generated by the algorithm. Given the correlation time of the noise and the total integration time, I would expect that the statistical scattering on the second moment of the noise is around 1.0%. The result of calculating the various moments, for an integration time step equals to 0.05 (the most favorable case for the present algorithm), is summarized in table 2. It is clear that the present algorithm generates a noise which has a distribution which is very different from a gaussian one: the second moment is some 25% larger than expected, whereas the higher moments are much smaller than what they should be. In fact, a much more reasonable value for $N$, such that the noise has the correct distribution (on this time scale) is $N \approx 2000$, which would, however, make the whole algorithm some factor 100 slower.

Table 1: Times taken by different algorithms

| Time step | Modified Heun | Present Algorithm |
|---|---|---|
| 0.1 | 11.17 | 11.00 |
| 0.05 | 22.13 | 14.53 |

Table 2: Noise moments generated by the different algorithms

| Quantity | Modified Heun | Present Algorithm | Theoretical |
|---|---|---|---|
| $\langle y \rangle$ | $-2.98 \, 10^{-3}$ | $-1.63 \, 10^{-3}$ | 0.0 |
| $\langle y^2 \rangle$ | 1.028 | 1.265 | 1.0 |
| $\langle y^3 \rangle$ | $-7.67 \, 10^{-3}$ | $-8.51 \, 10^{-2}$ | 0.0 |
| $\langle y^4 \rangle / (\langle y^2 \rangle^2)$ | 3.063 | 2.299 | 3.0 |
| $\langle y^6 \rangle / (\langle y^2 \rangle \langle y^4 \rangle)$ | 5.224 | 3.766 | 5.0 |

### 3.3 Quasi conservative systems

Another important class for which dedicated algorithms can be derived is given by the equations of motion

$$\dot{x} = v$$
$$\dot{v} = -\gamma v + F(x) + \xi(t) \tag{87}$$

where $\xi(t)$ is a random gaussian noise, with zero average and standard deviation

$$\langle \xi(t)\xi(s) \rangle = 2\gamma D \delta(t-s). \tag{88}$$

The above equation is commonly found in the liquid state literature, and several algorithms have been proposed, over the years, for its integration. The definitive word is perhaps summarized in [56], where two algorithms are proposed (see also references therein). The first algorithm integrates the above equation using the prescription

$$x(t+h) = x(t) + c_1 h v(t) + c_2 h^2 F(x(t)) + \eta_1 \tag{89}$$
$$v(t+h) = c_0 v(t) + c_1 h F(x(t)) + \eta_2, \tag{90}$$

where

$$c_0 = e^{-\gamma h} \tag{91}$$
$$c_1 = \frac{1-c_0}{\gamma h} \tag{92}$$
$$c_2 = \frac{1-c_1}{\gamma h} \tag{93}$$

and where $\eta_1$ and $\eta_2$ are two random gaussian variables with zero averages and moments

$$\langle \eta_1^2 \rangle = \frac{Dh}{\gamma}\left(2 - \frac{3 - 4e^{-\gamma h} + e^{-2\gamma h}}{\gamma h}\right) \tag{94}$$
$$\langle \eta_2^2 \rangle = D\left(1 - e^{-2\gamma h}\right) \tag{95}$$
$$\langle \eta_1 \eta_2 \rangle = \frac{D}{\gamma}\left(1 - e^{-\gamma h}\right)^2. \tag{96}$$

It is possible to check which is the equilibrium distribution reproduced by this algorithm. We know that the theoretical equilibrium distribution of Eq. 87 should be given by

$$P(x,v) = N \exp\left\{-\left[v^2/2 + V(x)\right]/D\right\} \tag{97}$$

where $N$ is a normalization constant and $V(x) = -\int F(x)dx$. Writing the distribution generated by the numerical scheme in the form

$$P(x,v) = N \exp\left\{-\left[v^2/2 + V(x) + hS(x,v)\right]/D\right\} \tag{98}$$

we have that $S(x,v)$ satisfies the differential equation

$$\frac{\partial^2 S(x,v)}{\partial v^2} - \frac{v}{D\gamma}\frac{\partial S(x,v)}{\partial x} - \left(\frac{F(x)}{D\gamma} - \frac{v}{D}\right)\frac{\partial S(x,v)}{\partial v} - \frac{v^2}{2\gamma D}F'(x) - \frac{1}{2\gamma}F'(x) = 0. \tag{99}$$

This means that this algorithms fails to reproduce the correct equilibrium distribution at $O(h)$ in the exponent. It is also possible, for the case when $V(x) = \omega^2 x^2/2$, to derive the equilibrium distribution at lower order in $h$, which reads

$$P(x,v) = N \exp\left\{-\left[v^2/2 + \omega^2 x^2/2\right]/D'\right\} \tag{100}$$

with

$$D' = \frac{D}{1 + \frac{\omega^2 h}{2\gamma}} \tag{101}$$

which means that when $\gamma$ goes to zero, the effective temperature simulated by the algorithm goes to zero (it must be said that in [56] it is acknowledged that the algorithm does not work well in this limit, although no formal proof is provided).

To overcome the problems with the case of small $\gamma$, in [56] a second algorithm is proposed, which reads

$$x(t+h) = x(t) + c_1 h v(t) + c_2 h^2 F(x(t)) + \eta_1 \tag{102}$$
$$v(t+h) = c_0 v(t) + (c_1 - c_2)hF(x(t)) + c_2 hF(x(t+h)) + \eta_2, \tag{103}$$

which indeed reproduces the correct equilibrium distribution at $o(h)$ (of course, there are corrections at $O(h^2)$ in the equilibrium distribution, but I will not write them here).

However, it is clear that both algorithms are fairly expensive in terms of CPU times: they require at least two random gaussian deviates and one or two evaluations of the force at each integration time step. Beside the Heun scheme and the full algorithm, which could be applied also to this case, it should be possible to find a more efficient algorithm for this specialized case.

The basic idea is that if we took $\gamma = 0$ in Eq. 87, I could integrate this equation using a simplectic scheme (see, for instance [57]). The scheme which is straightforwardly used in the presence of the noise would integrate Eq. 87

as

$$\tilde{x} = x(t) + \frac{h}{2}v(t) \tag{104}$$

$$v(t+h) = v(t) + hF(\tilde{x}) \tag{105}$$

$$x(t+h) = \tilde{x} + \frac{h}{2}v(t+h). \tag{106}$$

It is then possible to reintroduce both the dissipation ($-\gamma v$) and the noise, to obtain the scheme

$$\tilde{x} = x(t) + \frac{h}{2}v(t) \tag{107}$$

$$v(t+h) = c_2\left[c_1 v(t) + hF(\tilde{x}) + d_1\eta\right] \tag{108}$$

$$x(t+h) = \tilde{x} + \frac{h}{2}v(t+h), \tag{109}$$

where $\eta$ is a gaussian variable, with standard deviation one and average zero, and

$$c_1 = 1 - \frac{\gamma h}{2} \tag{110}$$

$$c_2 = \frac{1}{1 + \gamma h/2} \tag{111}$$

$$d_1 = \sqrt{2D\gamma h}. \tag{112}$$

This last algorithm does reproduce the correct equilibrium distribution (apart from terms $o(h^2)$, as the second algorithm proposed in [56]), but it only requires one gaussian deviate and one evaluation of the force. Furthermore, in the limit $\gamma \to 0$, by construction it conserves the energy of the system, at $o(h^3)$.

## 4  Conclusions

I have shown algorithms to integrate general stochastic differential equations, discussing connected problems. I would like to add that any simulation in stochastic dynamics requires a good and fast random number generator. Very many generators are available in the literature: personally, after trying a few, I settled for a couple of algorithms which satisfied my needs, and which have the good features that are fairly fast and portable to all platforms (they are written in high level languages). For the generation of flat distributions, I use the subtract and borrow algorithm, first proposed by Marsaglia (RCARRY), in the implementation by Luscher [58,59]. For the generation of gaussian or exponential deviates, I use the Ziggurath algorithm, by Marsaglia and Tsang [60], modified to use the output from the subtract and borrow routines. Let me finish noticing that no simulation can be better than the random noise used: so, extreme care should be given to the choice of the random noise generator. Most library random noise generators are reasonable for small simulations, but when a large number of random terms is needed, or the gaussianicity and true randomicity of the noise is required, my strong advice is to test very carefully the algorithm one is going to use. To this end, the benchmarks proposed in [61] are still the state of art.

## Acknowledgments

This work was partially support by EC under grant ERBCHRXCT930331